\documentclass[sigconf, 10pt]{acmart}
\usepackage{tikz}
\usepackage{amsmath}
\usepackage{filecontents}
\usepackage{xspace}
\usepackage{flushend}

\usepackage{graphicx}   
\usepackage{subcaption} 
\usepackage{soul}

\def\Section {\S}

\newcommand{\lxcc}{\texttt{LXCC}\xspace}
\newcommand{\carbonContainerS}{\textsc{Carbon Containers}\xspace}
\newcommand{\carbonContainer}{\textsc{Carbon Container}\xspace}

\AtBeginDocument{%
  }

\setcopyright{rightsretained}

\copyrightyear{2023}
\acmYear{2023}
\setcopyright{acmlicensed}\acmConference[SoCC '23]{ACM Symposium on
Cloud Computing}{October 30-November 1, 2023}{San Jose, CA, USA}
\acmBooktitle{ACM Symposium on Cloud Computing (SoCC '23), October
30-November 1, 2023, San Jose, CA, USA}
\acmPrice{15.00}
\acmDOI{10.1145/3620678.3624644}
\acmISBN{979-8-4007-0387-4/23/11}

\begin{document}
\sloppy

\date{}

\title[\textsc{Carbon Containers}]{\textsc{Carbon Containers}: A System-level Facility for Managing Application-level Carbon Emissions}

\author{John Thiede, Noman Bashir, David Irwin, Prashant Shenoy}
\affiliation{%
  \institution{University of Massachusetts Amherst}
  \country{}
}


\vspace{-5.5cm}



\begin{abstract}
To reduce their environmental impact, cloud datacenters' are increasingly focused on optimizing applications' carbon-efficiency, or work done per mass of carbon emitted. To facilitate such optimizations, we present \carbonContainerS, a simple system-level facility, which extends prior work on power containers, that automatically regulates applications' carbon emissions in response to variations in both their workload's intensity and their energy's carbon-intensity. Specifically, \carbonContainerS enable applications to specify a maximum carbon emissions rate (in g$\cdot$CO$_2$e/hr), and then transparently enforce this rate via a combination of vertical scaling, container migration, and suspend/resume while maximizing either energy-efficiency or performance.  

\carbonContainerS are especially useful for applications that i) must continue running even during high-carbon periods, and ii) execute in regions with few variations in carbon-intensity.  These low-variability regions also tend to have high average carbon-intensity, which increases the importance of regulating carbon emissions. We implement a \carbonContainer prototype by extending Linux Containers to incorporate the mechanisms above and evaluate it using real workload traces and carbon-intensity data from multiple regions. We compare \carbonContainerS with prior work that regulates carbon emissions by suspending/resuming applications during high/low carbon periods. We show that \carbonContainerS are more carbon-efficient and improve performance while maintaining similar carbon emissions. 
\vspace{-0.9cm}
\end{abstract}

\maketitle

\vspace{-0.2cm}
\section{Introduction}
\label{sec:intro}
The number and size of cloud datacenters is continuing to grow at a rapid pace to satisfy computing's increasing demand, which is being driven by a variety of new and computationally-intensive applications in artificial intelligence (AI) and machine learning (ML)~\cite{openai}.  This rapid growth in datacenter capacity will translate into rapid growth in energy consumption if improvements in computing's energy-efficiency do not keep pace with its capacity growth.  That is, every time datacenter capacity doubles, energy-efficiency must also double to keep energy consumption constant.  A recent report estimates datacenter capacity increased by 6$\times$ in the 2010s and is expected to increase by more in the 2020s~\cite{masanet}. Unfortunately, after decades of optimization, there are few remaining opportunities for further increasing energy-efficiency by a factor of 6$\times$ or more~\cite{google-pue}. 

The trends above have led to increasing concern about cloud computing's carbon emissions and impact on climate change moving forward.   As a result, there has been substantial recent work on optimizing applications' carbon-efficiency, or work done per mass of carbon emitted~\cite{ecovisor,wait-awhile,mccallum,enabling-socc21,dean-carbon,cdn1}.   Much of this work leverages variations in grid energy's carbon-intensity (in grams of carbon dioxide equivalent per kilowatt-hour or g$\cdot$CO$_2$e/kWh) to schedule computation when and where low-carbon energy is available, e.g., by migrating load to low-carbon regions or deferring load to low-carbon periods.  Grid energy's carbon-intensity varies spatially because each region has its own mix of energy sources, which have different carbon intensities. For example, solar, wind, hydro, geothermal, and nuclear have zero marginal carbon emissions, while natural gas-powered generators tend to have fewer carbon emissions than coal-powered generators.  Likewise, energy's carbon-intensity also varies temporally because the mix of energy sources (with different carbon intensities) the grid uses in each region changes over time due to changes in both demand and weather.  Recent work has developed carbon-aware policies for i) temporal workload shifting by suspending jobs when grid energy's carbon-intensity exceeds a configurable threshold subject to deadline constraints~\cite{wait-awhile} and ii) spatial workload shifting by routing web requests to regions with excess solar energy subject to latency constraints~\cite{cdn1}. This work has shown significant potential for reducing carbon emissions in and across regions with widely variable carbon-intensity. 

Unfortunately, prior work on temporal shifting does not apply to either applications that must execute continuously or in most high-carbon regions, as these regions have few variations in carbon-intensity, while spatial shifting for stateful workloads often incurs prohibitive overheads.  To address the problem, this paper presents \carbonContainerS, a simple system-level facility that regulates the carbon emissions of individual applications in response to variations in both their workload's intensity and their energy's carbon-intensity.  \carbonContainerS extend the notion of Power Containers~\cite{power-containers}, a previously proposed OS facility for fine-grained power and energy management in servers, and combines it with resource deflation and migration techniques \cite{sharma2019resource,hotspot} to regulate an application's carbon emissions.  

Specifically, \carbonContainerS enable applications to specify a configurable maximum carbon emissions rate (in g$\cdot$CO$_2$e/hr), and then transparently enforce this rate via a combination of vertical scaling, container migration, and suspend/resume, while maximizing either performance or energy-efficiency.   That is, instead of either suspending jobs (or migrating them to a lower-carbon region) when carbon-intensity increases, \carbonContainerS deflates their resource allocation by vertically scaling them down to ensure they do not exceed their maximum rate.  If vertical scaling is either insufficient or too inefficient, \carbonContainerS enforce the target carbon emissions by automatically migrating to a server with a lower energy and carbon footprint, i.e., a smaller server.  \carbonContainerS only suspend themselves when energy's carbon-intensity is so high that vertical scaling and migration cannot satisfy the carbon target. 

Importantly, beyond setting the target carbon emissions rate, \carbonContainerS' operation is entirely transparent to applications, unlike recent work on virtualizing the energy system, which exposes carbon-intensity dynamics to applications and makes them responsible for optimizing their own carbon-efficiency~\cite{ecovisor}.  Our hypothesis is that \carbonContainerS provides a general and flexible tool for transparently managing application carbon emissions in response to variations in workload- and carbon-intensity. In evaluating our hypothesis, we make the following contributions. 

\noindent {\bf Carbon- and Workload-Intensity Data Analysis}.  We analyze grid carbon-intensity and cloud workloads in production traces to understand how they vary.  We show that, while grid carbon-intensity typically has few variations (on the order of hours-to-days), job resource usage, and thus energy consumption, in production workloads varies widely (on the order of minutes-to-hours).  We also show that high-carbon regions, where managing carbon is most important, have few variations in carbon-intensity. Our analysis motivates that adapting applications to changes in their workload-intensity is just as, if not more, important as adapting to changes in energy's carbon-intensity in managing carbon emissions. 

\noindent {\bf \carbonContainerS Design}. We present the design of \carbonContainerS, which builds on Power Containers~\cite{power-containers} by transparently enforcing a configurable maximum carbon emissions rate for applications via a combination of vertical scaling, migration, and suspend/resume.  We develop two enforcement policies for \carbonContainerS that prioritize energy-efficiency or performance. The former minimizes an application's energy consumption while minimally throttling it, while the latter always operates close to the carbon emissions target regardless of its energy-efficiency.  

\noindent {\bf Implementation and Evaluation}. We implement a Linux \carbonContainerS (\lxcc) prototype by extending Linux Containers (LXC), and evaluate it on CloudLab and in simulation using production job and carbon-intensity traces.   We compare \lxcc with a recent approach that controls carbon emissions by suspending/resuming applications during high/low carbon periods~\cite{wait-awhile}, and show that \carbonContainerS are significantly more carbon-efficient in enabling higher performance for only a small increase in emissions. We have publicly released \carbonContainerS under a permissive open-source license.\footnote{\url{https://github.com/carbonfirst/CarbonContainers}}

\vspace{-0.1cm}
\section{Motivation and Background}
\label{sec:background}
In this section, we motivate \carbonContainerS by analyzing real-world data on grid energy's carbon-intensity (\Section\ref{sec:grid}), cloud datacenters' workload-intensity (\Section\ref{sec:cloud}), and their impact on both energy- and carbon-efficiency (\Section\ref{sec:carbon}). 

\begin{figure}
    \centering
    \includegraphics[width=\linewidth]{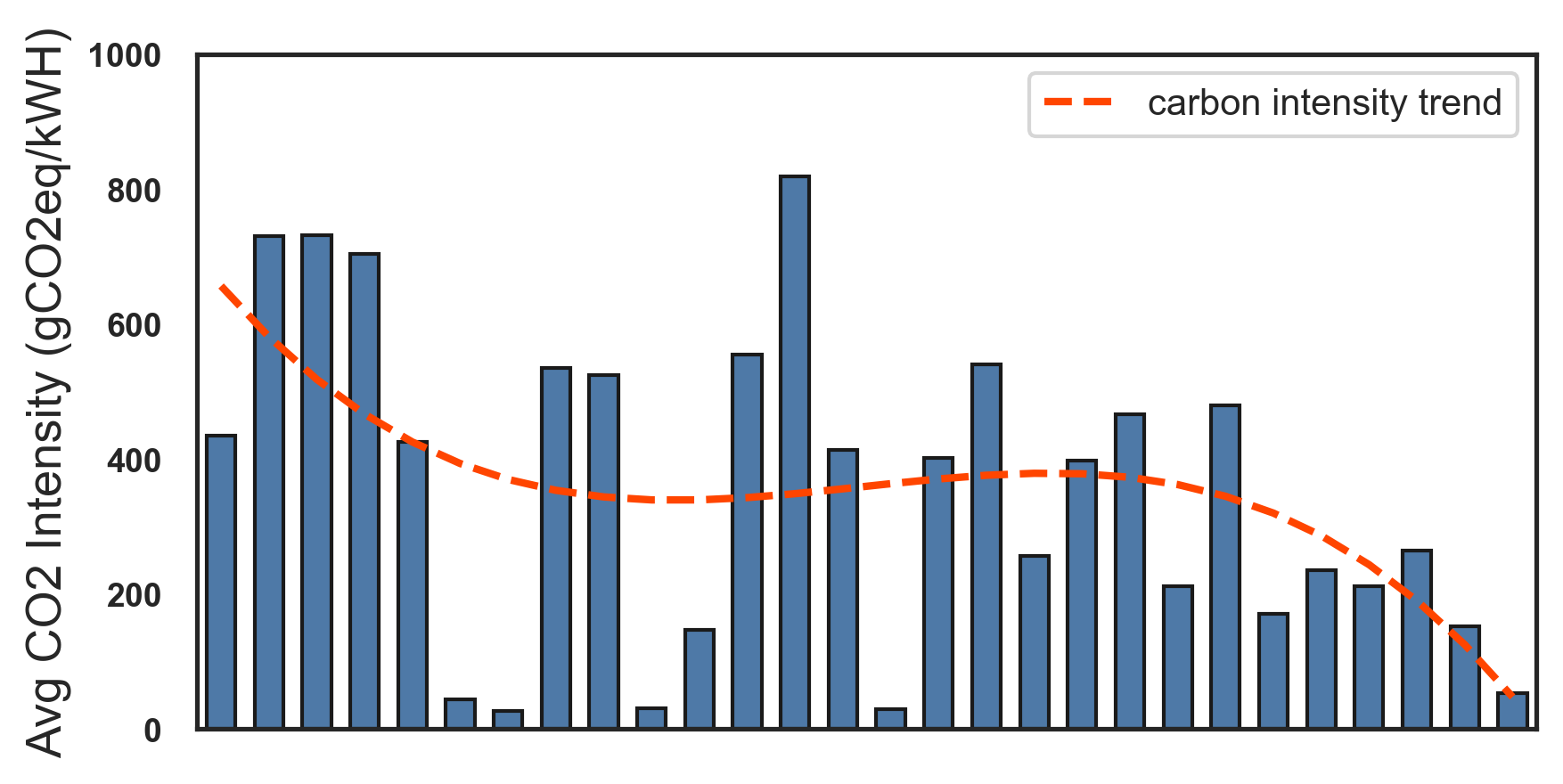}
    \\
    \vspace{-0.1cm}
    \includegraphics[width=0.99\linewidth]{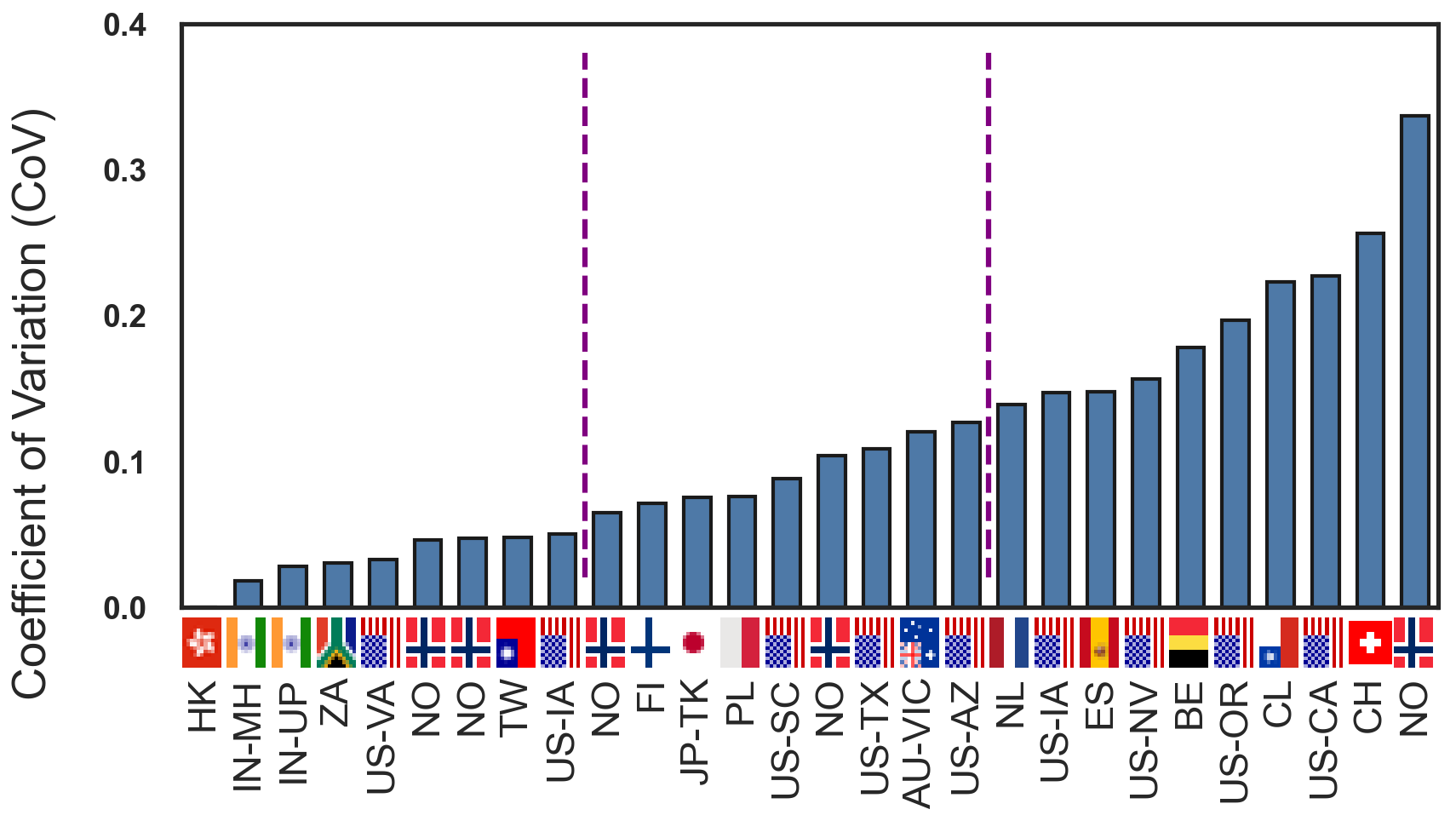}
    \vspace{-0.3cm}
    \caption{\emph{The average carbon-intensity (top) and Coefficient of Variation (CoV) (bottom) for many regions worldwide.  The x-axis is ordered by increasing CoV.  The locations with high average carbon-intensity generally have 
    a low CoV, albeit with some notable exceptions.}}
    \label{fig:carbon-characteristics}
    \vspace{-0.6cm}
\end{figure}

\vspace{-0.15cm}
\subsection{Grid Energy's Carbon-Intensity}
\label{sec:grid}

As mentioned in \S\ref{sec:intro}, grid energy's average carbon-intensity in g$\cdot$CO$_2$e/kWh varies over time based on the changing mix of generators (with different carbon-intensities) it uses to satisfy a variable demand.  In addition, different regions have widely different carbon-intensity characteristics in terms of both their average magnitude and variance.  For example, Figure~\ref{fig:carbon-characteristics} shows both the average carbon-intensity (top) and Coefficient of Variation (CoV) (bottom) for 27 regions worldwide.  This data comes from electricityMap~\cite{electricity-map}, a carbon information service that estimates per-region carbon emissions based on public data about the type and output of generators used in each region over time.  ElectricityMap reports the average carbon-intensity every hour. Since the set of active generators and their output changes relatively slowly (based primarily on day/night cycles), grid energy's carbon-intensity does not change significantly within an hour.  We use this data to compute the annual daily carbon-intensity CoV (based on hourly readings).  The CoV is the ratio of a dataset's standard deviation to its mean.  Thus, a CoV at or near $0$ indicates nearly constant data. The $x$-axis for both graphs in Figure~\ref{fig:carbon-characteristics} is in order of increasing CoV.

The graph demonstrates multiple key points.  Most importantly,  there is a wide difference between the lowest carbon region and the highest carbon region---by more than 500$\times$---as some regions have large quantities of zero-carbon energy sources, e.g., hydro, nuclear, geothermal, solar, wind, etc., while others have almost none.  Clearly, managing carbon emissions in the higher carbon regions is much more important, and has a much bigger impact, than managing them in lower carbon regions. For example, reducing energy usage by only 5\% in a region where carbon-intensity is 800g$\cdot$CO$_2$/kWh lowers emissions more than reducing energy usage by 100\% (which is impossible) in a region where carbon-intensity is only 30g$\cdot$CO$_2$/kWh (assuming carbon-intensity is constant). 

\begin{figure}
    \centering
    \includegraphics[width=0.96\linewidth]{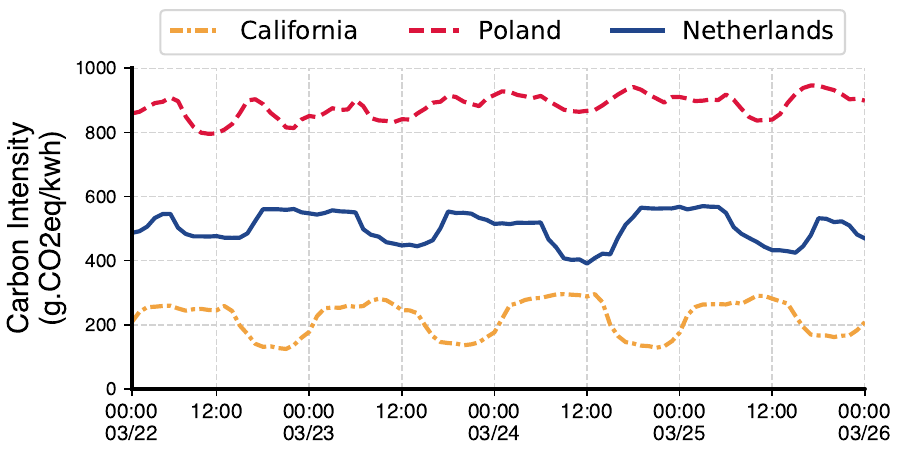}
        \vspace{-0.4cm}
    \caption{\emph{Energy's carbon-intensity for representative regions over a 96-period with low (Poland), medium (Netherlands), and high (California) CoV (see Figure~\ref{fig:carbon-characteristics}).}}
        \vspace{-0.75cm}
    \label{fig:carbon-example}
\end{figure}

Notably, there is also a wide difference between the regions with the lowest and highest CoV.   Many CoVs are quite low, as nearly one-third of the regions have a daily hourly CoV below $0.05$, as indicated by the vertical dotted lines, which divides the regions into thirds.  A CoV$<$$0.05$ indicates that energy's carbon-intensity is nearly constant.  Further, CoVs for the middle third of regions range from only $0.05$ to $0.15$, which, while higher, is sill relatively low.  Only the last couple of regions in our dataset have much higher CoVs ($0.15$ and $0.35$) due to large penetrations of renewable energy.  Figure~\ref{fig:carbon-example} illustrates the differences in CoV for $3$ different carbon regions over 48 hours: one in each third of Figure~\ref{fig:carbon-characteristics}(bottom).   The graph shows minimal variations for the regions in the lowest and middle third of CoV (Poland and the Netherlands), and more variation for the region in the highest third third of CoV (California), largely due to high solar penetration.  

Our illustrative example also shows that the regions with less variation tend to have a higher average carbon-intensity.   Figure~\ref{fig:carbon-characteristics}(top) shows that this is generally true across all 27 regions with a few notable exceptions. The figure plots the average carbon-intensity of each region in the same order as in Figure~\ref{fig:carbon-characteristics}(bottom). As shown, the average carbon-intensity trend generally decreases as the CoV increases, except for some regions mixed in with very low carbon-intensity.  These regions have a low CoV and low average carbon-intensity primarily due to the presence of large quantities of nuclear and hydro energy.  In all other cases, the decrease in carbon-intensity is due to increasing solar and wind energy, which increases the CoV.   Overall, the highest-third of regions in terms of CoV has an average carbon-intensity of 189g$\cdot$CO$_2$/kWh; the middle-third has 344g$\cdot$CO$_2$/kW; and the lowest-third has 551g$\cdot$CO$_2$/kW (or nearly 2$\times$ more than high CoV regions). 

Importantly, carbon-aware scheduling policies that suspend/resume applications when carbon-intensity is high/low~\cite{wait-awhile} are only effective when carbon-intensity varies widely.  Unfortunately, our analysis shows this only happens in regions with low average carbon-intensity. Thus, while suspend/resume policies may yield a significant local percentage reduction in carbon emissions, their absolute reduction is quite small. In addition, the lack of variations in carbon-intensity across many regions means that dynamically migrating jobs to the lowest carbon region is ineffective for jobs with any memory and disk state. The migration overhead is high, and the carbon-intensity across different regions rarely intersects.   As Figure~\ref{fig:carbon-example} illustrates, low-carbon regions tend to always have lower carbon-intensity than high-carbon regions.  Thus, even if we ignore migration overhead, there are few times when moving from one region to another substantially lowers carbon emissions. 

\vspace{-0.1cm}
\subsection{Datacenters' Workload-Intensity}
\label{sec:cloud}

\begin{figure}
    \centering
    \includegraphics[width=\linewidth]{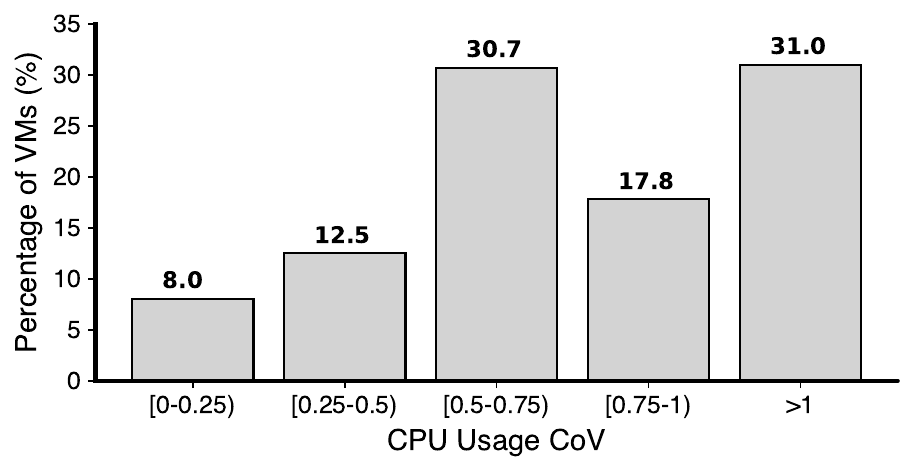}
            \vspace{-0.8cm}
    \caption{\emph{Percentage of VMs from a random 1000 VM in the Azure trace with different ranges of CoV.}}
            \vspace{-0.45cm}
    \label{fig:azure-graph}
\end{figure}

We also analyzed jobs' workload-intensity from a production job trace.  In this case, we analyze a public trace released by Azure that provides the minimum, maximum, and average CPU utilization and memory allocation for $\sim$2.7 million production virtual machines (VMs) every 5 minutes over 30 days.  The Azure trace has a size of 235GB and contains $\sim$1.9 billion readings~\cite{azure-data}.   There are two primary takeaways from our trace analysis that motivate our work.  

\emph{High Workload Variations}. Most importantly, VM CPU utilization exhibits potentially wide variations on the order of minutes to hours.   While a few VMs exhibit constant resource usage, most exhibit some variance.  For example, Figure~\ref{fig:azure-graph} shows that, from a random sample of 1000 VMs in the Azure trace, only 8\% have a CoV below $0.25$.  In this case, we compute the CoV over 5-minute intervals, rather than the 1-hour intervals in the carbon-intensity data.  Thus, even the low CoVs suggest more variation than the carbon-intensity traces.  In addition, 30\% of VMs have CoV greater than $1$ which indicates extremely high variance (i.e., where standard deviation exceeds the mean), and over 50\% of VMs have CoV greater than $0.4$.  Overall, the variations in workload-intensity are much larger than those in energy's carbon-intensity.  

\emph{Low Resource Utilization}. The second important takeaway from our workload-intensity analysis is that average CPU utilization across VMs is typically low with more than 43\% of the VMs having less than 10\% utilization.  In general, low utilization is highly energy-inefficient. Since servers are not energy-proportional~\cite{energy-proportional}, their most efficient operating point is at 100\% utilization, as this amortizes their baseload power across the most amount of computation.  Baseload power is non-trivial and can be as high as 50\% of a server's peak power.  As a result, migrating jobs across servers as their utilization changes can have a substantial effect on their energy-efficiency, and thus also their carbon-efficiency. 

\vspace{-0.3cm}
\subsection{Impact on Carbon-Efficiency}
\label{sec:carbon}

Our analysis above motivates our design for \carbonContainerS, which regulates an application's carbon emissions in response to variations in both carbon- and workload-intensity using a combination of vertical scaling, migration, and suspend/resume.   As we show, \carbonContainerS primarily {\em adapt to changes in an application's workload-intensity, as it varies much more than carbon-intensity.}  

Importantly, when an application's workload-intensity changes on a server, so does its energy-efficiency and thus carbon-efficiency, as carbon emissions are simply the product of an application's energy consumption and its energy's carbon-intensity.   Specifically, when workload-intensity decreases, energy-efficiency also decreases since servers are not energy-proportional.  At some point, migrating to a smaller server, i.e., with fewer cores and less memory, can increase energy-efficiency and thus carbon-efficiency, since it reduces baseload power and amortizes it over the same computation.  As we discuss, \carbonContainerS' enforcement policy leverages this insight to satisfy its carbon target, while minimally throttling resources and maximizing energy-efficiency. 

As summarized below, \carbonContainerS address multiple problems with existing techniques for leveraging temporal and spatial variations in energy's carbon-intensity using suspend/resume scheduling and wide-area migration. 

\emph{Ineffective in high carbon regions.}  Suspend/resume scheduling policies that suspend jobs when carbon emissions are high and resume them when low are only effective in regions where energy's carbon-intensity varies widely~\cite{wait-awhile}.   That is, these techniques are only effective if energy's carbon-intensity is periodically low. Yet, as we show in \Section\ref{sec:grid}, energy's carbon-intensity does not vary widely in many regions, largely due to a low penetration of intermittent solar and wind energy sources, which cause most of the variations. As a result, the regions with high carbon emissions (by a wide margin), where managing carbon emissions is the most critical, tend also to be the ones where suspend/resume scheduling policies are the least effective.  In contrast, \carbonContainerS can enforce an arbitrary carbon emissions rate regardless of the variations in grid energy's carbon-intensity, and thus can be effective even in high-carbon regions with few variations.   As mentioned above, \carbonContainerS mostly adapt to frequent and significant changes in a job's workload-intensity rather than energy's carbon-intensity.

\emph{High performance penalty.} Even in regions where carbon-intensity varies widely, it typically follows a diurnal pattern with significant changes occurring on the order of hours.  Thus, reducing carbon emissions using suspend/resume policies often requires delaying jobs by many hours---from night to day. This high performance penalty is likely undesirable for many shorter batch jobs, and prohibitive for interactive jobs, which require an immediate response.  Thus, by rate-limiting rather than suspending jobs, \carbonContainerS lowers the performance penalty due to high carbon periods compared to suspend/resume policies: jobs always keep running but at a lower performance level.    

\emph{High migration overhead.}  One way to prevent delaying jobs when carbon-intensity increases is to migrate them to lower carbon regions~\cite{followsun}. While dynamically migrating (or routing) small web/inference requests to low-carbon regions (as part of geo-replicated services) is possible~\cite{cdn1}, migrating stateful jobs with non-trivial memory or disk state over the wide area incurs a high performance and energy overhead~\cite{wan-migration}.  This overhead limits the applicability and benefit of carbon-aware spatial workload shifting. Further, cross-region migration is only useful between regions with highly variable carbon-intensity that is also out-of-phase, where low-carbon periods are not aligned.  However, as we show in \Section\ref{sec:grid}, there are few such regions.  In contrast, \carbonContainerS shows that job migration can be an effective tool for managing carbon emissions \emph{within a datacenter}, even when all servers share the same energy and thus carbon-intensity. While all servers share the same energy, datacenters include different types of servers.  Since the energy-efficiency of applications with time-varying demand is different on different types of servers, their carbon-efficiency is also different.

\section{\carbonContainerS Design}
\label{sec:design}
In this section, we first present \carbonContainerS' architecture (\Section\ref{sec:architecture}), which builds on Linux Containers (LXC), and then discuss its carbon enforcement policy (\Section\ref{sec:policy}). 

\vspace{-0.2cm}
\subsection{System Architecture}
\label{sec:architecture}

\carbonContainerS enable users to configure a maximum (or target) carbon emissions rate (in g$\cdot$CO$_2$e/hr), which they transparently enforce via a policy that combines vertical scaling, container migration, and suspend/resume. We assume the target carbon rate is set based on an exogenous policy that captures applications' tradeoff between performance and carbon emissions. Of course, setting a lower target carbon rate may decrease performance, e.g., when a high carbon-intensity period aligns with high utilization. Alternatively, a policy may choose to set a higher target carbon rate to service a critical application, and accept much higher carbon emissions. There is no free lunch in optimizing carbon; only tradeoffs.   As we discuss, \carbonContainerS' goal is to maximize an application's energy-efficiency while minimally throttling performance, as workload- and carbon-intensity vary, subject to its target carbon emission rate. 

\begin{figure}
    \centering
    \includegraphics[width=\linewidth]{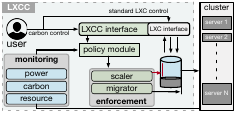}
    \vspace{-0.75cm}
    \caption{\emph{High-level architecture for \carbonContainerS, including its monitoring, policy, and enforcement modules on a cluster with servers of varying sizes.}}
    \vspace{-0.65cm}
    \label{fig:design}
\end{figure}

Figure~\ref{fig:design} depicts the \carbonContainerS architecture, which builds on existing container functions.  While we build on LXC, the architecture is general and could interface equally well with other container implementations.  Since our prototype builds on LXC, we refer to it as Linux \carbonContainerS or \lxcc.  \carbonContainerS operate from the perspective of a cloud user, rather than provider, and thus make decisions locally without considering a cloud platform's carbon emissions.  While providers should consider their whole infrastructure in optimizing carbon emissions, cloud users can only control their applications. The \carbonContainerS core is a policy module that runs as a background daemon and i) registers newly created containers, ii) monitors containers' workload-, power-, and carbon-intensity, and iii) controls vertical scaling, container migration, and suspend/resume functions to enforce each container's carbon target.  We discuss these basic functions and policies in \S\ref{sec:policy}. 

\vspace{-0.15cm}
\subsubsection{Carbon Container Interface.}  \lxcc's policy module includes a programmatic interface for registering new containers, setting their maximum carbon target, and configuring their enforcement policy.  The policy module also includes a configuration file that captures information on the types of servers available for migration, the information necessary for requesting them, i.e., a cloud API key, and information on monitoring per-container power usage and energy's carbon-intensity.  As we discuss in \S\ref{sec:implementation}, we built an {\tt lxcc} command-line program that wraps the {\tt lxc} command-line tool and interfaces with the policy module's programmatic interface.  Our command-line tool passes most commands through to lxc, retaining the same interface and options as {\tt lxc}, but adds new {\tt lxcc}-specific options, e.g., for setting the carbon target, and also registering and de-registering containers with the policy module when they are created and destroyed.  We made the \lxcc interface as similar as possible to LXC. 

\vspace{-0.15cm}
\subsubsection{Monitoring Subsystem}  The policy module includes a monitoring subsystem that monitors applications' resource, power, and carbon usage in real time, as discussed below.  

\noindent {\bf Resource monitoring}. \lxcc monitors per-core utilization, memory usage, and a range of other hardware performance counters, to determine if a container is under-utilizing its resources or potentially being throttled. Specifically, \lxcc estimates resource utilization on a scale from 0-100\%, such that 100\% utilization indicates a container that has been throttled, i.e., could use additional resources, while any utilization below 100\% is not throttled.  We use a performance model to normalize this utilization relative to a baseline server to make it comparable across different servers.  This utilization is also averaged across the cores assigned to a container.  

We assume a family of homogeneous server instances as in cloud platforms with resources that are fixed multiples of each other, and scale the estimated utilization across different servers linearly based on this multiple.  That is, these servers have the same hardware architecture, but with different resource allocations. This enables \lxcc to use a simple linear performance model to estimate resource usage on other servers.  For example, we assume a container running at 40\% utilization on the baseline server would run at 20\% utilization on a server 2$\times$ larger and 80\% utilization on a server $0.5$$\times$ smaller. For simplicity, \lxcc optimistically assumes that a throttled container operating at 100\% utilization would operate at 50\% utilization on a server 2$\times$ as large, e.g., it can utilize more cores.  If this assumption is wrong, then the enforcement policies will self-correct, as the actual utilization will be less than expected, which will trigger the enforcement policy. Note that a container that is highly utilizing a large server instance may have a normalized utilization greater than 100\%, indicating that it would be throttled on the baseline server.  Since \lxcc's goal is to throttle containers as little as possible, its enforcement policy will only throttle them once they are at or near the carbon target.  

\lxcc also uses the resource usage above to infer a container's real-time power usage, as discussed below.  In general, to support more heterogeneous servers, i.e., with different hardware architectures, \lxcc could use more sophisticated performance models, e.g., using machine learning, that infer the resource usage on other servers from the resource usage on the server it is currently executing on. However, adding such support is future work.  \lxcc monitors resource usage at a high resolution, e.g., every five minutes.

\noindent {\bf Power monitoring.} \lxcc requires the ability to monitor fine-grained power on a per-container basis, as in prior work on Power Containers~\cite{power-containers}.   In particular, \lxcc supports model-based power monitoring based on performance counters. Users can supply their own configurable model for each server, or use external power monitoring software, such as PowerAPI~\cite{power-api}, that includes such models. Thus, \lxcc can leverage the substantial prior work on developing power models~\cite{power-api}. These power models must be configured for \lxcc based on the particular set of servers it runs on. They include a base power component, which is the power at idle and does not change with utilization, and a marginal power component, which dynamically varies with utilization. 

As prior work shows, CPU utilization remains the dominant component of marginal power consumption, as it still has by far the widest dynamic power range~\cite{power-api}.   We also show this experimentally in \Section\ref{sec:implementation} for the servers used in our evaluation. When determining whether to migrate a container, \lxcc combines the performance model above with its power model to estimate what a container's utilization and power would be on other servers, and thus requires power models for other candidate servers as well. \lxcc monitors power usage at a high resolution, e.g., every five minutes.  

\noindent {\bf Carbon monitoring.} \lxcc interfaces with electricityMap's API~\cite{electricity-map} to monitor the carbon-intensity of its servers' energy based on their operating region.    \lxcc enables users to specify the region in a configuration file.  As mentioned earlier, carbon-intensity changes only every hour, so the policy module only updates it once per hour.   Given average power consumption $p(t)$ over its monitoring interval $\Delta$ and energy's carbon-intensity $c(t)$, the policy module also monitors a container's overall carbon emissions rate $C(t)$=$p(t) \times c(t)$.   As we discuss in \S\ref{sec:policy}, the enforcement policy responds if $C(t)$ is at or close to the carbon target rate $C_{target}$.    

\begin{figure}
    \centering
    \includegraphics[width=\linewidth]{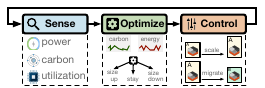}
    \vspace{-0.75cm}
    \caption{\emph{\carbonContainerS' workflow for monitoring energy usage and carbon-intensity to make container-level carbon management decisions.}}
    \vspace{-0.9cm}
    \label{fig:workflow} 
\end{figure}

\vspace{-0.2cm}
\subsubsection{Enforcement Mechanisms.} \lxcc's enforcement policy uses a combination of 3 mechanisms --- vertical scaling, container migration, and suspend/resume --- discussed below to ensure containers remain at or below their carbon target.

\noindent {\bf Vertical scaling}.  \lxcc sets resource quotas to cap the maximum power usage as in recent work~\cite{google-osdi20}.   Specifically, \lxcc uses LXC cgroups to control the number of cores a container can use and (optionally) the time slice for each core.   By capping resource usage, vertical scaling caps the power usage and carbon emissions rate for a given carbon-intensity.  As we discuss in \S\ref{sec:policy}, \lxcc adjusts containers' resource quota in response to changes in energy's carbon-intensity to ensure they never exceed their target carbon emissions rate.   

Vertical scaling has two potential drawbacks when enforcing a carbon target.  Most importantly, since servers typically have a low dynamic power range, vertical scaling alone may not satisfy a carbon target if carbon-intensity significantly increases, as happens in the evening in regions with a high solar penetration.  In such cases, even if the resource quota is set to $0\%$, a server's baseload power may cause it to exceed its carbon target.  In addition, vertically scaling down a container's quota decreases its energy-efficiency, since it amortizes the server's baseload power over less computation. 

\noindent {\bf Container migration.}  Container migration addresses both drawbacks of vertical scaling.  In essence, migrating a container to smaller and larger servers effectively extends \lxcc's dynamic power range.   In general, smaller servers, e.g., with fewer cores and memory, also have a proportionately lower baseload power, and thus are more energy-efficient; they can also be just as performant as larger servers for workloads that cannot fully utilize a larger server.   As discussed in \S\ref{sec:policy}, \lxcc migrates to larger servers, if the carbon emissions allow, to prevent throttling a container, and migrates to smaller servers to enforce the carbon target once it becomes more energy-efficient than further vertical scaling down.  \lxcc includes a configurable table of servers available for migration, e.g., as provided by cloud platforms, where each container locally determines where to migrate based on its own policy.

As we discuss in \S\ref{sec:implementation}, LXC supports both checkpoint/restore and live migration~\cite{xen-migration}, although there are some restrictions on the container configuration.   Live migration is transparent and incurs little downtime, while a checkpoint/restore migration requires pausing the container, transferring its memory state to the destination server, and then restoring it.  While both approaches can maintain active TCP connections as long as the downtime is less than the TCP timeout, a checkpoint/restore migration incurs a performance overhead as the application cannot execute during the migration. 

\noindent {\bf Suspend/Resume}. \lxcc can also suspend a container, which idles its server and drops its marginal power usage to $0$.  However, since servers always consume baseload power, suspending containers is infinitely energy-inefficient, as they perform no useful work but still consume substantial energy.  \lxcc's enforcement policy only suspends a container when it cannot operate below the carbon target by migrating to the smallest (most energy-efficient) server and vertically scaling it down to minimize the baseload power that is wasted when suspended. The baseload power of the smallest server dictates a lower bound on \lxcc's power usage.  Thus, there are scenarios where it is impossible for \lxcc to enforce its carbon target if the carbon-intensity increases too much. 

\vspace{-0.2cm}
\subsection{Carbon Enforcement Policy}
\label{sec:policy}
Given a target carbon emissions rate for a container, \lxcc transparently executes an enforcement policy that combines the mechanisms above to ensure the container does not exceed its rate.    \lxcc's enforcement policy has two variants. The default policy ensures a container does not exceed the target carbon emissions rate, minimizing energy consumption without throttling the container, i.e., by never operating at 100\% utilization.    We call this policy the \emph{energy-efficiency policy}, since it prioritizes energy-efficiency.   In contrast, users may also configure an alternative \emph{performance policy}, which ensures that a container's emissions rate always remains within some threshold of the target rate.   Thus, under the performance policy, a container may run on a large, power-intensive server at low usage, as long as it remains below its carbon target, which is highly inefficient. The performance policy is useful for providing a container reserve capacity to handle any sudden load bursts with low latency. 

In effect, the energy-efficiency policy variant enforces the target carbon rate, while also minimizing overall carbon emissions, while the performance policy variant operates at or near the target.   Both policy variants minimize throttling the container subject to the target carbon rate, i.e., they only throttle when necessary to enforce the carbon target.    Below, we first discuss general aspects of both enforcement policy variants, and then discuss their specific differences. 

\vspace{-0.25cm}
\subsubsection{General Enforcement Policy}
Both policy variants continuously compare the current carbon emissions $C_i(t)$ of a container on its current server $i$ to its carbon target $C_{target}$.   If $C_i(t)$ comes within some configurable threshold $\epsilon$, \lxcc triggers an enforcement mechanism.  As described above, $C_i(t)$ is a function of a container's resource utilization (and thus power usage) and energy's carbon-intensity.  The value of $\epsilon$ is configurable and presents a tradeoff.   If the value is near $0$, i.e., actions are only enforced when at the target, then the container i) may periodically exceed $C_{target}$ since enforcement actions have some delay, and ii) may cause thrashing that triggers unnecessary enforcement actions, i.e., migrations. As $\epsilon$'s value increases, the policy diverges more from the strict target, but lessens the overhead due to thrashing. 

The first enforcement mechanism is to vertically scale a container down until $C(t)$ is not within the threshold, as vertical scaling has lower overhead than migration.  In parallel, the policy also estimates, based on the power model of the next smallest server $j$, the carbon emissions rate $C_j(t)$.   As the policy vertically scales down a container, if the carbon emissions rate $C_j(t)$ on the next smallest server ever drops below $C_i(j)$ \emph{and} the smaller server throttles the application less than vertically scaling down the larger server, the policy triggers a migration of the container to the smaller server.  

To illustrate the decision of when to migrate versus continue vertically scaling, consider the following example with a ``big'' server that has 2$\times$ the resource capacity of a ``small'' server, where we assume the big server has a baseload power of 100W and peak power of 200W, while the small server has a baseload power of 50W and peak power of 100W. If the big server is throttled by 50\%, i.e., capped at 50\% utilization, it would consume 150W, but have the same performance capacity as a small non-throttled server consuming 100W.  At this point, assuming the container is fully using its 50\% allocation on the big server, the policy would migrate to the smaller server as it provides the same performance for less energy. Note that if \lxcc requests to provision a server from the list for migration, and it is not available, then \lxcc removes the server and re-evaluates the policy.

At some point, if both a container's workload- and carbon-intensity increase too much, the policy will migrate the application to the smallest server such that further migrations are impossible, and the container is fully throttled due to vertical scaling.  At this point, the policy suspends the container until carbon-intensity decreases to a point where the container can be vertically scaled up and is not throttled. 

In addition to scaling containers down when $C_i(t)$ approaches $C_{target}$, the policy may also scale containers up if their resource utilization increases, and they are below $C_{target}$.  Similar to above, in this case, the policy vertically scales containers up until they reach $C_{target}$ or they have access to the server's entire resource capacity.  If a container is fully utilizing a server's resource capacity, and it is still below $C_{target}$, then the server is throttled, and the policy will migrate the container to the next largest server (as long as doing so would not exceed $C_{target}$).  

\vspace{-0.25cm}
\subsubsection{Energy-efficiency Variant}
The energy-efficiency policy variant extends the general policy above by simply migrating containers to smaller servers if they are not fully utilizing their current server.  In this case, the migration decision is essentially the same as the one above, but is triggered instead based on a lack of server utilization rather than forced vertical scaling due to being near $C_{target}$.  That is, in the example above, if a container is only utilizing the big server 50\% or less, rather than being vertically scaled down to 50\%, the decision is the same:  migrating to the smaller server will be more energy-efficient and carbon-efficient, and doing so will not throttle the container.  Thus, the energy-efficiency variant will migrate the container down in this case. Notably, the energy-efficiency variant still ensures that containers are never throttled if they are below $C_{target}$.  That is, the policy does not simply maximize energy-efficiency, as doing so would require always executing a container on the smallest most energy-efficient server regardless of throttling. 

\vspace{-0.25cm}
\subsubsection{Performance Variant}
Unlike the energy-efficiency variant above, the performance policy variant \emph{does not} migrate containers to smaller servers when they are below $C_{target}$ and are not fully utilizing their current server.   Instead, the performance policy attempts to vertically scale up and migrate containers to larger servers to be within $\epsilon$ of the carbon target \emph{regardless of a container's utilization}.  As a result, the performance policy is less energy-efficient, as it may run an idle container on a large server $i$ if energy's carbon-intensity is low, as long as the container's carbon emissions rate $C_i(t)$ remains below $C_{target}$.   Thus, the performance variant uses its excess carbon to maintain reserve capacity to handle unexpected bursts in resource usage.   Since many jobs have a low average usage interspersed with large bursts of utilization, the performance variant tends to incur less migrations and overhead from migrating containers to smaller servers after a burst of resource and power usage.

\vspace{-0.1cm}
\section{Implementation}
\label{sec:implementation}
We implemented a \carbonContainerS prototype in python 3.7+ using a microservice approach consisting of a collection of coordinating services that run as background daemon processes and communicate via gRPCs.  Our implementation uses Linux Containers (LXC 3.0.3)~\cite{linux-containers} and CRIU v3.7 (Checkpoint/Restore in Userspace)~\cite{criu} for migration.   CRIU supports container checkpoint/restore (or stop-and-copy) and live migration, although the implementations are highly sensitive to the container configuration and its set of running processes.  For our experiments, we configured containers such that these mechanisms would work. In particular, our containers include a stock 64-bit Ubuntu Xenial image.  Our prototype includes i) a front-end command-line tool for creating, configuring, and destroying \carbonContainerS, ii) a monitoring service for resource, power, and carbon usage, and iii) a policy module that receives data from the monitoring service and triggers enforcement mechanisms based on the policy in \S\ref{sec:policy}.  We discuss each service's implementation below, and then present prototype microbenchmarks.

\begin{figure}
    \centering
    \includegraphics[width=0.99\linewidth]{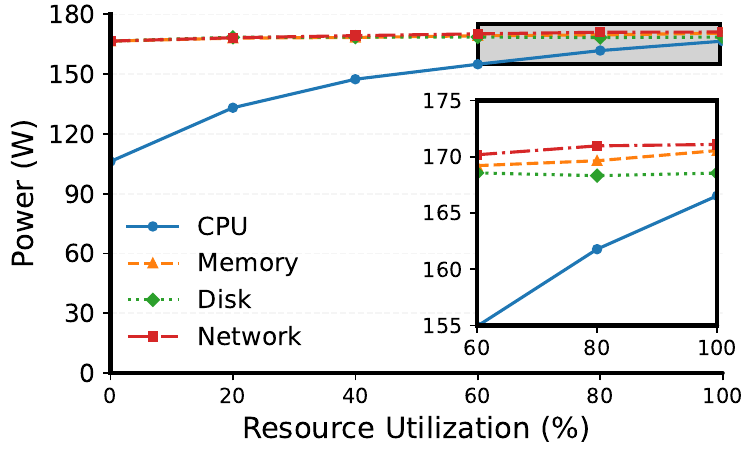}
        \vspace{-0.4cm}
    \caption{\emph{Measured power usage relative to resource utilization levels. CPU usage is set at 100\% to isolate the effect of memory, network, and disk from CPU.}}
        \vspace{-0.6cm}
    \label{fig:micro-power-scale}
\end{figure}

\vspace{0.05cm}
\noindent
\textbf{LXCC services.} \lxcc uses a configuration file that includes information on the available server types, their power models, API keys for cloud platforms and carbon information services, and ssh keys for accessing other servers.  Our prototype uses a simple power model that includes a server's baseload and peak power such that power usage increases linearly from the baseload to the peak power based on server utilization. These simple models were highly accurate when calibrated to our servers.  In particular, Figure~\ref{fig:micro-power-scale} shows the power usage of our local server (a Dell PowerEdge R430), as a function of the resource utilization of its CPU, memory, disk, and network resources.  Here, we used the stress-ng workload emulator to utilize each resource in isolation at a specific percentage.  Since utilizing any resource also increases CPU utilization, for all resources except  CPU, we conducted the experiments with the CPU at 100\% utilization.  

Figure~\ref{fig:micro-power-scale} demonstrates both that i) the memory, disk, and network have little dynamic power range, since there is little difference in power at 100\% utilization for any resource (see inset) and ii) the relationship between CPU utilization and power consumption is roughly linear. We experimented with other power models, including fitting a cubic polynomial and training machine learning models using performance counters, but found that these models were not significantly more accurate than a simple linear model. In general, the relationship between resource usage and power is server-specific, and depends on the dynamic ranges of a server's components.  As a result, the simple linear power model above may not apply to other servers. However, \carbonContainerS is agnostic to the precise power model, and can support arbitrarily complex power models that are a function of any values available to the monitoring service, which include a wide range of performance counters.

We intend our prototype to operate on cloud platforms, where it requests new servers dynamically, when migrating a container to a smaller or larger server.  In this case, the policy module issues a request to a cloud API to provision a server before migrating to it.  We assume these cloud servers boot an image with the \lxcc services running, and are accessible via the same ssh keys.   Here, we assume a one-to-one ratio between \carbonContainerS and cloud servers (which may run as VMs).  In addition, \lxcc can also operate from a static set of servers; our experiments on CloudLab use this approach, since CloudLab does not provide a programmatic API for dynamically provisioning servers. 

\begin{figure}
    \centering
    \includegraphics[width=\linewidth]{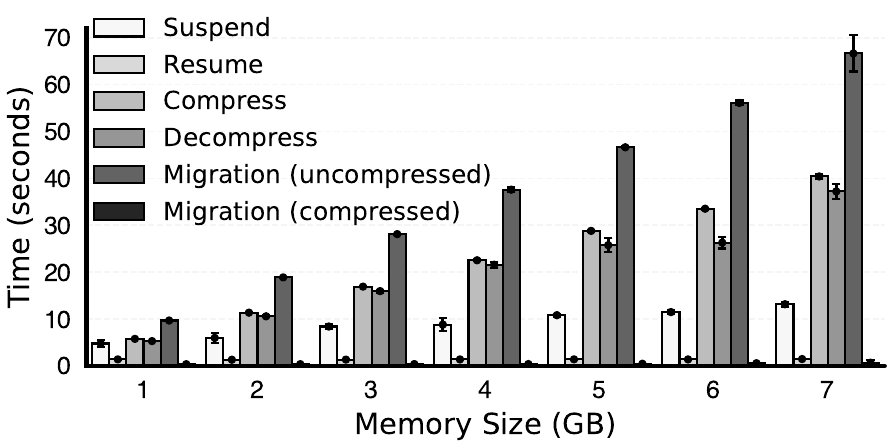}
        \vspace{-0.8cm}
    \caption{\emph{The time required to suspend/resume, compress/decompress, and migrate \lxcc containers as a function of memory footprint. }}
        \vspace{-0.65cm}
    \label{fig:micro-memory}
\end{figure}

\vspace{0.05cm}
\noindent {\bf Command-line tool}. The {\tt lxcc} command-line program wraps the normal {\tt lxc} tool and provides minimal additional functionality.  The {\tt lxcc} tool enables users to view current information on a container's resource, power, and carbon usage by fetching data from the monitoring service.  The tool also enables users to create, configure, and destroy \carbonContainerS.   When creating a container, the tool registers the container with the monitoring and policy modules.  The tool also enables configuring containers by setting their target carbon rate $C_{target}$, $\epsilon$ threshold, and policy variant (i.e., energy-efficiency versus performance policy). 

\vspace{0.05cm}
\noindent {\bf Monitoring module}.  The monitoring module tracks energy's carbon-intensity via electrictyMap's API.  The module maps processes to specific containers and tracks their resource utilization.  The service uses this utilization as input to the power models above to track estimated power usage and carbon emissions rate, both on the current server and the other available server types.  The monitoring module includes an API that enables other services to query its data.   The monitoring module also writes resource usage, power, and carbon data to disk for historical analysis. 

\vspace{0.05cm}
\noindent {\bf Policy module}. The policy module polls the monitoring service for each container's carbon emissions rate and resource usage every interval, e.g., 5 minutes by default, and implements the enforcement policy from \S\ref{sec:policy}.  Our prototype implements vertical scaling using Linux \texttt{cgroups}, by controlling the number of cores a container can use. As mentioned above, our prototype uses CRIU for migration.  When performing a stop-and-copy migration, the policy module checkpoints the container, compresses its filesystem, configuration, and checkpoint files, and transfers them to the destination server. The policy module at the destination service receives the archive, decompresses it, relocates the container filesystem and configuration to LXC's directory (\texttt{/var/lib/lxc}), and restores the container from the CRIU snapshot.

\vspace{-0.6cm}
\subsection{Microbenchmarks}

We next benchmark the performance of various sub-tasks that \carbonContainerS perform. 

\noindent {\bf Migration overhead}. Figure~\ref{fig:micro-memory} shows the time required to suspend/resume, compress/decompress, and migrate \lxcc containers as their memory footprint increases on a CloudLab server (d430) with 32 CPU cores and 62 GB of memory.  In this case, the migration is from a d430 server to a d820 server.  The results are the average of 10 experiments, where the error bars represent the standard deviation.  We separate the time to suspend/resume and compress/decompress, and also show the migration time with both compressed and uncompressed memory images. Of course, the migration overhead depends on the size of a container's memory and disk state.  Here, we migrate a container's memory-resident working set, which varies, along with a small root disk.

Our results offer two key insights.  First, the time to migrate the uncompressed image is the dominant time, and roughly equal to compressing, migrating, and decompressing the image.  The time to migrate the compressed image is negligible given the high compression ratio. Notably, this migration time is significantly less than the time needed to suspend/resume. Second, the time for all operations is roughly linear with memory size, although with different slopes.  Suspend/resume and compress/decompress scale more gracefully, i.e., have smaller slopes, than migrating an uncompressed image.  Nevertheless, the experiments also show that even for relatively high memory footprints, e.g., 7GB, the migration time for a stop-and-copy migration is still less than 2 minutes.  Of course, a live migration incurs no downtime, although it does incur some energy cost from requiring two servers to operate at the same time. 

\begin{figure}
    \centering
    \includegraphics[width=\linewidth]{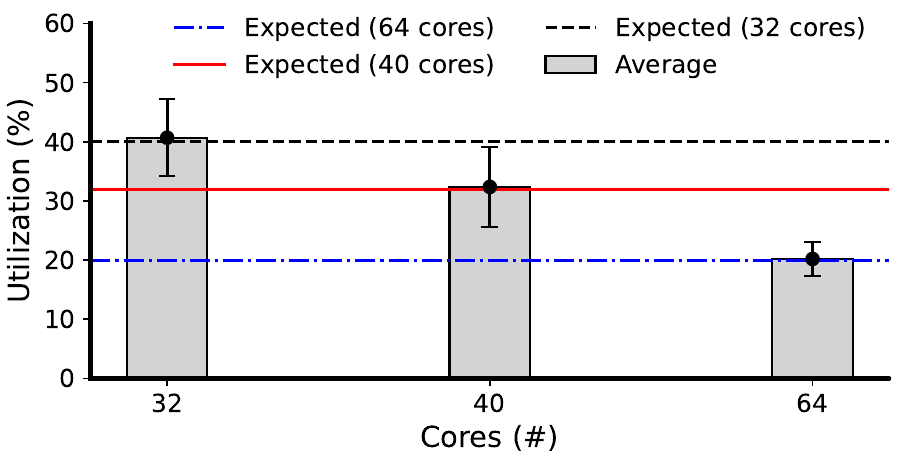}
        \vspace{-0.8cm}
    \caption{\emph{Effect of migrating to a server of a different size on resource utilization.}}
        \vspace{-0.6cm}
    \label{fig:micro-size}
\end{figure}

\vspace{0.03cm}
\noindent {\bf Server performance comparison}.  Our prototype uses a simple power model that assumes server performance and power usage scales linearly with resource capacity.  Figure~\ref{fig:micro-size} validates this assumption for a compute-intensive job that operates at 40\% utilization on our baseline server with 32 cores. We migrate the workload to servers with 40 and 64 cores, and verify that the utilization changes proportionately, as expected.  Figure~\ref{fig:micro-size} shows the actual utilization on each server, and the expected utilization based on the core ratio. 

\begin{figure}
\centering
\includegraphics[width=\linewidth]{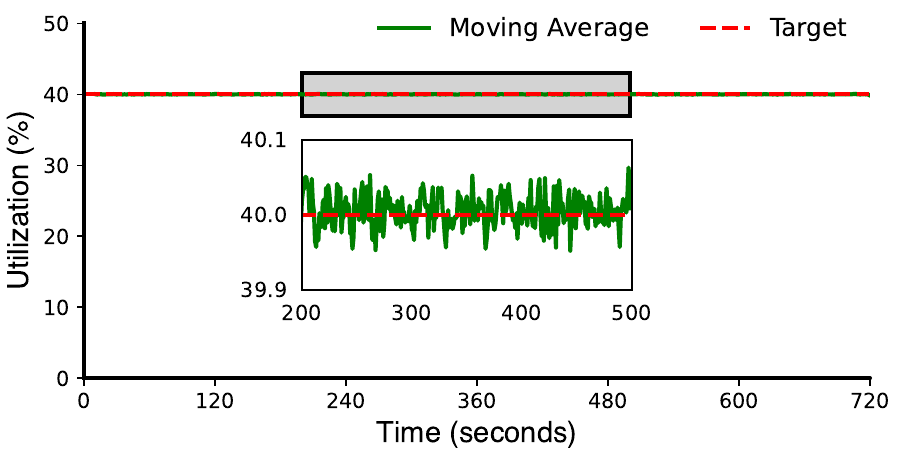}
\vspace{-0.8cm}
\caption{\emph{Efficacy of our prototype in replaying workload traces. While the instantaneous utilization varies, the moving average is within < 1\% of the target usage.}}
\vspace{-0.5cm}
\label{fig:micro-replay}
\end{figure}

\vspace{0.03cm}
\noindent {\bf Workload emulator}. Finally, we use a workload emulator (stress-ng) to replay utilization traces on servers.  We run a microbenchmark to verify that stress-ng can maintain a configurable utilization.  Figure~\ref{fig:micro-replay} shows the result of using stress-ng on a 64-core server above at 40\% utilization.  The graph shows that stress-ng maintains a utilization within $<$1\% of the target 40\% utilization. Here, we monitor CPU every 5 seconds and report a moving average over 1 minute.

\section{Experimental Evaluation}
\label{sec:evaluation}
In this section, we evaluate the performance of \carbonContainerS. We first present experiments that demonstrate the ability of our \carbonContainerS prototype, \lxcc, to enforce an arbitrary carbon emissions target. We then evaluate \carbonContainerS' enforcement policy in simulation at large-scale across a wide range of workload characteristics and carbon-intensity scenarios, and compare them with a recent suspend/resume scheduling approach~\cite{wait-awhile}. 

\subsection{Evaluation Setup}
Below, we describe our \carbonContainerS evaluation setup, various baselines, and specific evaluation metrics.

\vspace{-0.1cm}
\subsubsection{Traces}

We use two types of traces in our evaluation: resource usage traces from production cloud workloads and carbon-intensity traces for different geographical regions in the world. For resource usage, we use a Microsoft Azure trace~\cite{azure-data, azure-data-paper} that provides the minimum, maximum, and average CPU and memory usage information for $\sim$2.7 million VMs every 5 minutes over a 30-day period. We sample 1000 VMs at random for our large-scale analysis in simulation. For carbon-intensity information, we use the average carbon-intensity information from electricityMaps~\cite{electricity-map}. The traces provide hourly carbon-intensity values for all the regions in the world. Since our enforcement policy performs differently based on the variance in carbon-intensity, we select representative regions that have high (Netherlands) and medium (California) variations in their carbon-intensity, as discussed in \S\ref{sec:background} and shown in Figure~\ref{fig:carbon-example}. Ultimately, \carbonContainerS benefits depend on both applications' pattern of workload demands and carbon-intensity. If neither workload demand nor carbon-intensity vary, there is little room for reducing carbon emissions without degrading performance.

\vspace{-0.15cm}
\subsubsection{Baselines}

We compare \carbonContainerS with three baselines: carbon-agnostic, suspend/resume, and \carbonContainerS with vertical scaling without migration. 

For the carbon-agnostic approach, we assume a job runs on a baseline server without any vertical scaling or migration.  For suspend/resume scheduling, we assume a job also runs on a baseline server without any vertical scaling or migration.  In this case, the scheduler suspends a job when its rate of carbon emissions falls below the target carbon rate, and resumes it once it rises above.  Finally, we also implement a variant of \carbonContainerS that uses vertical scaling and suspend/resume but does not migrate containers.  That is, this policy will attempt to satisfy the carbon target by vertically scaling down, but if it cannot it suspends the container rather than migrating.  This is essentially a resource-aware version of suspend/resume scheduling.  When simulating \carbonContainerS, we assume jobs start on the same baseline server as above, but can migrate to one of five servers in the same family that are 4$\times$, 2$\times$, $0.5$$\times$, and $0.25$$\times$ the resource capacity.  We model these capacities after a family of general-purpose servers on a public cloud platform, specifically Amazon Web Services.  We assume the baseload and peak power of these servers is in proportion to their resource capacity, and that our baseline server has a baseload power of 100W and peak power of 200W with power usage between the base and peak scaling proportionate to utilization. 

\vspace{-0.2cm}
\subsubsection{Metrics}

We focus our evaluation on quantifying the average carbon emissions rate (in g$\cdot$CO$_2$e/hour) and the percentage an application is throttled, which represents its performance degradation. The throttling percentage is normalized relative to our baseline server, such that 10\% throttling on average represents a job that would have utilized a server with 110\% of the capacity of our baseline server.  The goal is to have both low average carbon emissions rate (at or below the target) and a low throttling percentage. 

\vspace{-0.1cm}
\subsection{Prototype Evaluation}

\begin{figure}
    \centering
    \includegraphics[width=0.96\linewidth]{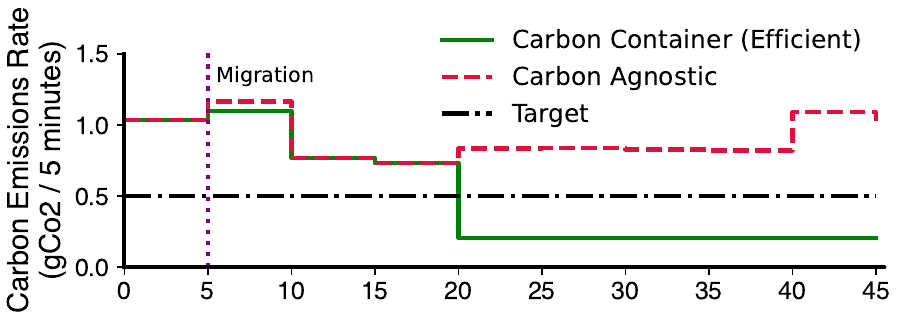}\vspace{-0.1cm} \\
    \includegraphics[width=0.96\linewidth]{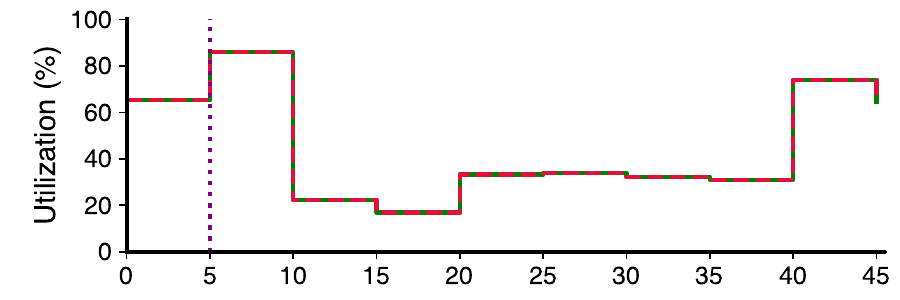} \vspace{-0.1cm} \\
    \includegraphics[width=0.96\linewidth]{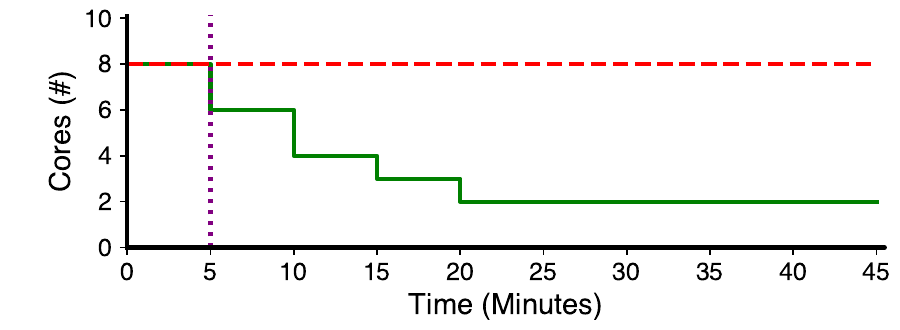}
    \vspace{-0.45cm}
    \caption{\emph{Illustration of our \carbonContainerS prototype. Since this trace is less than an hour duration, the carbon intensity is steady at 300.91 gCO$_2$/kWh.} } 
    \vspace{-0.6cm}
    \label{fig:timeseries_eff_highCoV}
\end{figure}

We first evaluate our \carbonContainerS prototype to demonstrate its salient features.  Figure~\ref{fig:timeseries_eff_highCoV} shows a time-series of our prototype running a job with variable workload-intensity over a nearly hour-long period.  We use our stress-ng workload emulator to replay the job within \carbonContainerS. The top graph shows the target carbon rate, as well as the average carbon emissions for our \carbonContainer and for a carbon-agnostic policy.  For this example, we use the energy-efficiency policy for \carbonContainerS.  

\begin{figure}[t]
    \centering
    \includegraphics[width=\linewidth]{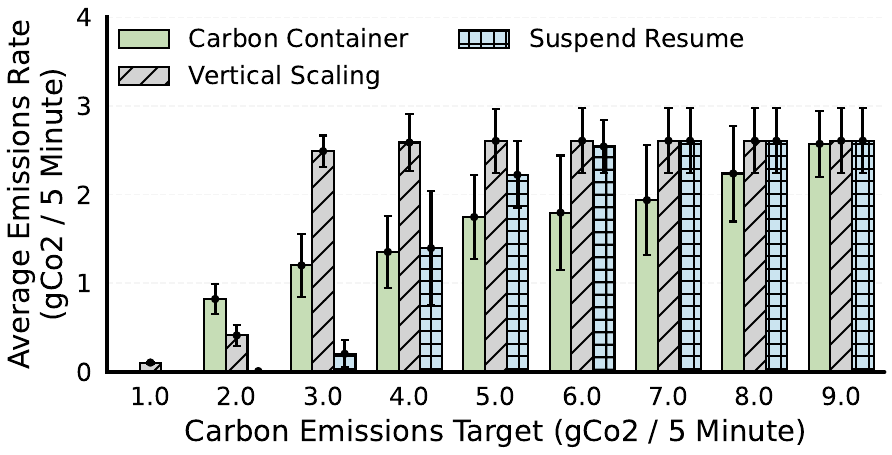}
    \vspace{-0.75cm}
    \caption{\emph{Average carbon emissions rate for \carbonContainerS and other baseline approaches in a region with highly variable carbon-intensity.}}
    \vspace{-0.35cm}
    \label{fig:cc_vs_others_carbon}
\end{figure}

\begin{figure}[t]
    \centering
    \includegraphics[width=\linewidth]{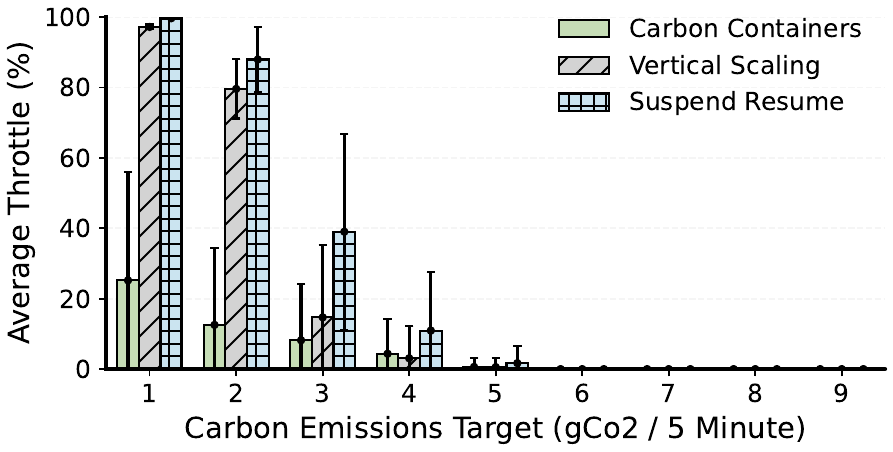}
    \vspace{-0.75cm}
    \caption{\emph{Average throttling for \carbonContainerS and other baseline approaches in a region with highly variable carbon-intensity (companion to Figure~\ref{fig:cc_vs_others_carbon})}}
    \vspace{-0.6cm}
    \label{fig:cc_vs_others_performance}
\end{figure}

The top graph shows that \carbonContainerS starts above the target but then recognizes this and migrates to a smaller server to get below the target.  In contrast, the carbon-agnostic approach remains above the target for the entire period. The middle graph shows the utilization of the container, which increases at the beginning of the trace but then decreases in the middle and then increases again at the end; both the \carbonContainer and the carbon-agnostic approach yield the same utilization, as they replay the same trace.  The bottom graph then shows the number of cores utilized by the container.  At the start, the \carbonContainer attempts to vertically scale down the container to reduce the carbon emissions before determining it must migrate to a smaller server to get emissions below the target. Here, the destination machine is a \texttt{pc3000} server (2 CPU cores), while the original server was a \texttt{d710} (8 CPU cores). This is annotated in each graph, and the results can be seen in the top graph as a significant reduction in the average carbon emissions rate. The prototype graph above demonstrates the basic functions of our \carbonContainerS prototype. 

\subsection{Large-scale Evaluation}

We next perform a larger-scale evaluation over more jobs and more regions. Note that our simulation experiments include the overhead from migration from our testbed.  Thus, we expect individual \carbonContainerS performance to follow our experiments.  In a production datacenter, performance may improve due to higher-capacity networking infrastructure.  In these experiments, we select a random sample of 1000 jobs from the Azure trace, and simulate their performance with \carbonContainerS.  We report averages across the jobs, as well as standard deviation using error bars. 

Figure~\ref{fig:cc_vs_others_carbon} shows the average carbon emissions rate of each approach at varying target carbon rates for our region with highly variable carbon-intensity, alongside the carbon emissions under a carbon-agnostic policy. We can see that \carbonContainerS manages to maintain a carbon emissions rate below the given target, even for small targets. That said, the other policies also operate below the carbon target.

\begin{figure}[t]
  \centering
  \includegraphics[width=\linewidth]{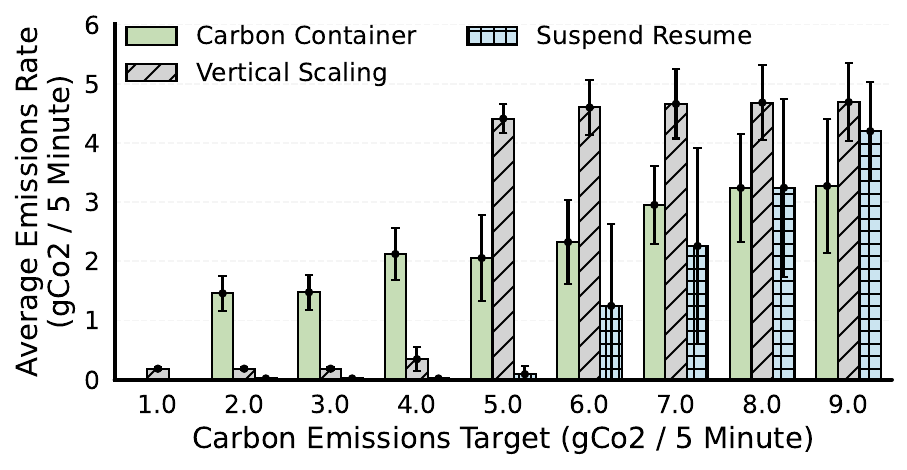}
  \vspace{-0.85cm}
  \caption{\emph{Average carbon emissions rate for \carbonContainerS and other baseline approaches in a region with medium variable carbon-intensity.}}
  \vspace{-0.5cm}
  \label{fig:cc_vs_others_carbon_med}
\end{figure}

\begin{figure}[t]
  \centering
  \includegraphics[width=\linewidth]{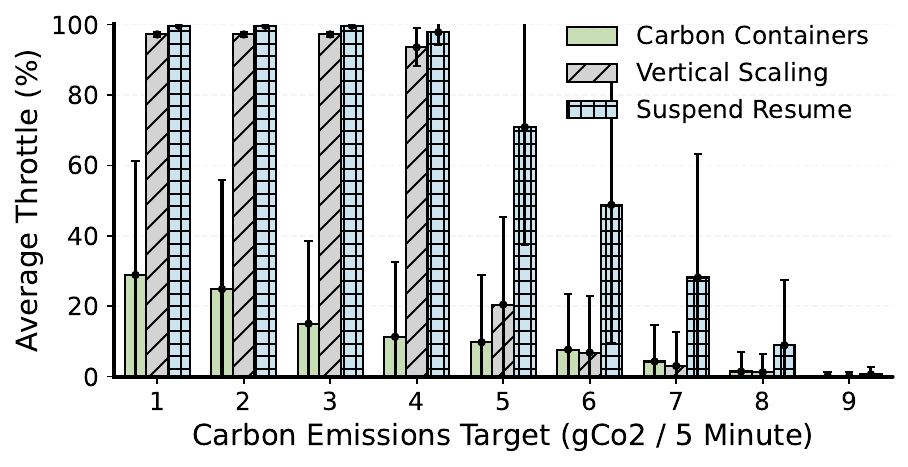}
  \vspace{-0.85cm}
  \caption{\emph{Average throttling for \carbonContainerS and other baseline approaches in a region with medium variable carbon-intensity (companion to Figure~\ref{fig:cc_vs_others_carbon_med}).}}
  \vspace{-0.35cm}
  \label{fig:cc_vs_others_performance_med}
\end{figure}

However, the carbon rate for the suspend-resume policy is misleading for low target values. Carbon savings alone fails to capture the advantage that migration and vertical scaling have over the other policies, especially suspend-resume. In particular, when a job is suspended, no forward progress is being made, and as such the suspend-resume approach substantially increases the time needed to finish a job. In this figure, many of the low carbon targets result in small emissions averages because the suspend-resume policy spends significant amounts of time not running. Vertical scaling reduces this penalty by throttling resources before forcing a full stop. This throttling also has an impact on the performance, based on the magnitude of resource reduction and the time spent at reduced resource levels. Suspension can be re-contextualized in terms of throttling by defining a suspension as a period of 100\% magnitude throttling. In this case, such vertical scaling naturally has a bound on the potential savings at 100\% resource reduction.   

\begin{figure}[t]
    \centering
    \includegraphics[width=\linewidth]{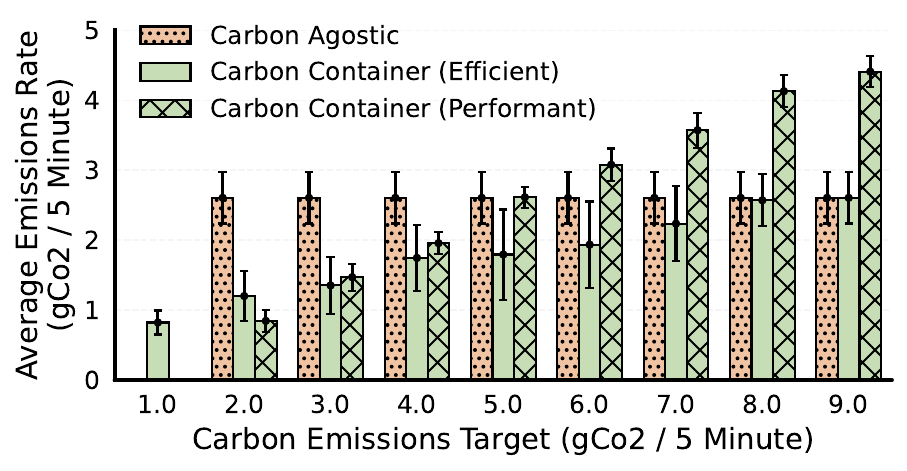}
      \vspace{-0.85cm}
    \caption{\emph{Average carbon emissions rate for the energy-efficiency and performance policy variants in a region with highly variable carbon-intensity.}}
    \vspace{-0.5cm}
    \label{fig:sim_eff_vs_perf_carbon_high}
\end{figure}

\begin{figure}[t]
    \centering
    \includegraphics[width=\linewidth]{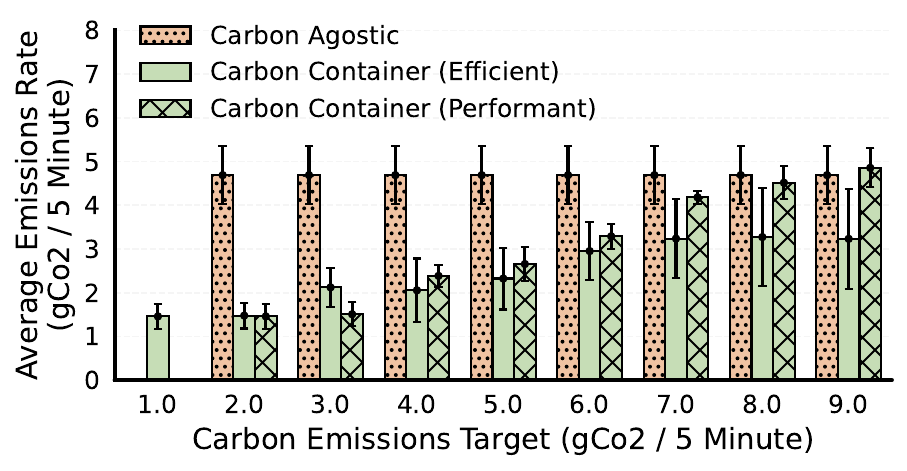}
      \vspace{-0.85cm}
    \caption{\emph{Average carbon emissions rate for the energy-efficiency and performance policy variants in a region with medium variability carbon-intensity.}}
    \vspace{-0.6cm}
    \label{fig:sim_eff_vs_perf_carbon_med}
\end{figure}

Figure \ref{fig:cc_vs_others_performance} then compares the performance throttling experienced by the jobs while operating under the given policies for the same experiment as Figure~\ref{fig:cc_vs_others_carbon}.  The suspend-resume approach naturally experiences the highest degree of throttling, as its only mechanism for avoiding exceeding a carbon threshold is to completely stop until the carbon-intensity decreases. Vertical scaling experiences less throttling due to the reasons stated above, while \carbonContainerS experiences the least amount of throttling by a significant margin. Due to \carbonContainerS' migration policy, its effective energy scaling range becomes much larger than using vertical scaling in isolation. By moving to smaller servers, the jobs can effectively reduce their minimum baseload energy requirements.  Due to this flexibility, \carbonContainerS rarely needs to fully suspend execution at any point. Migration is also highly effective because applications in cloud traces have high variances, as shown in Figure~\ref{fig:azure-graph}.  Thus, migrating to a smaller more energy-efficient server during a low-intensity period yields significant benefits.

Figures~\ref{fig:cc_vs_others_carbon_med} and \ref{fig:cc_vs_others_performance_med} show the same analysis for a region with medium variations. A similar pattern emerges as in our first region: for many of the lower end targets, suspend-resume fails as it waits indefinitely for a carbon-intensity reduction that never comes. \carbonContainerS themselves have a lower carbon bound that they cannot completely satisfy, but this limit is defined by the size of the smallest available server instead of a limit inherent to the region. In this case, for low-end targets, \carbonContainerS with migration experience some overhead that increases its carbon emissions relative to vertical scaling (although still operating below the target carbon emissions rate), but this comes with a substantial decrease in throttling. 

\vspace{-0.2cm}
\subsubsection{Energy-Efficiency vs Performance Policy Variants}

In \S\ref{sec:design}, we describe two variations of our carbon enforcement policy: an energy-efficiency and performance variant. As mentioned, we anticipate that aggressively optimizing for energy-efficiency may not be suitable for all use cases, as some applications and users may not be looking to minimize their carbon emissions, but rather maximize their performance while satisfying a carbon rate limit. Such applications would desire a policy that more aggressively scales up to larger servers to avoid throttling time and be better prepared to handle large bursts of demand. The  performance policy variant aims to accommodate these use cases. As such, we evaluate the two implementations of our policy against each other, and against a carbon-agnostic policy. 

\begin{figure}[t]
    \centering
    \includegraphics[width=\linewidth]{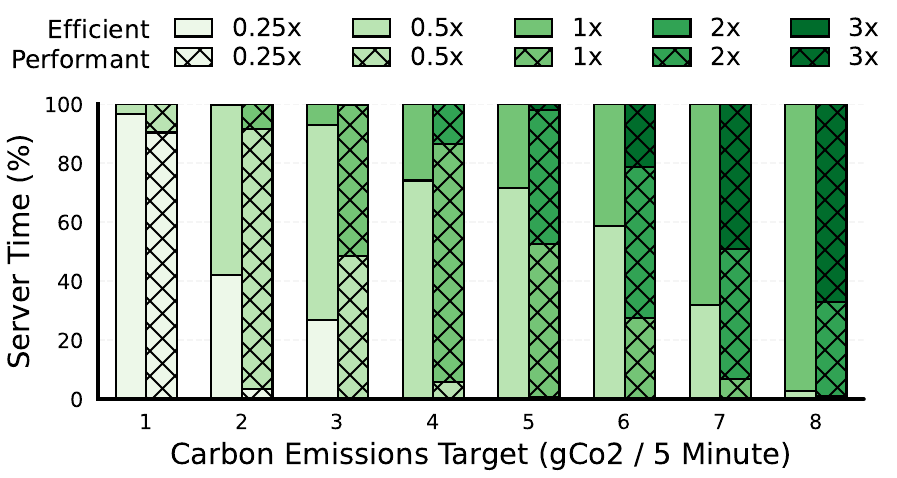}
      \vspace{-0.85cm}
    \caption{\emph{Percentage of time spent on different size servers by the performance and energy-efficiency policies in a high carbon variation region.}}
    \vspace{-0.9cm}
    \label{fig:comparison}
\end{figure}

Specifically, Figures~\ref{fig:sim_eff_vs_perf_carbon_high} and \ref{fig:sim_eff_vs_perf_carbon_med} compare the carbon emissions of each policy for our high carbon and medium carbon variation region.  These figures demonstrate how these different policies manage carbon emissions. As the carbon target increases, the performance policy variant is able to spend more time running on larger machines, resulting in more carbon emissions but also higher performance.  Figure~\ref{fig:comparison} then captures the difference in performance potential where the x-axis is again the carbon target, while the y-axis is the percentage of time spent on different size servers.  In particular, the figure shows that the performance policy spends a much larger fraction of time executing \carbonContainerS on larger, less energy-efficient servers.  However, note that both policy variants still satisfy the carbon target. 

\vspace{-0.1cm}
\section{Related Work}
\label{sec:related}
\carbonContainerS is related to a range of prior work on power, resource, and carbon management on cloud platforms, which we discuss below.  Most importantly, \carbonContainerS differs from much of this prior work in that it focuses on providing a mechanism for enforcing a carbon target without dictating how it might be used.  We envision that \carbonContainerS could be used in a wide variety of higher-level systems, such as carbon-aware cluster schedulers for batch/service jobs, serverless functions, etc.

\vspace{0.07cm}
\noindent {\bf Power management.}  \carbonContainerS are directly inspired by prior work on Power Containers~\cite{power-containers}.  Indeed, \carbonContainerS essentially extend Power Containers by enforcing a target carbon rate that includes not only power consumption but also energy's carbon-intensity.   We also designed \carbonContainerS with cloud platforms in mind by enabling them to self-migrate between different types of servers as their utilization (and thus energy-efficiency) changes.  \carbonContainerS sets power caps by placing quotas on resource usage, which is a common technique used by many prior systems~\cite{google-osdi20,capping-asplos,rosing-isca}.  However, prior work generally caps power to prevent server clusters from exceeding the power delivery infrastructure's maximum power rating.  In our case, \carbonContainerS cap power to prevent exceeding a target carbon emissions rate. 

\vspace{0.07cm}
\noindent {\bf Resource management}.  There has also been a variety of work that uses containers to adjust resource usage on cloud platforms, often in response to price changes.  For example, HotSpot migrates containers to different servers in response to changes in spot prices~\cite{hotspot}.   However, HotSpot focuses on maximizing an application's cost-efficiency, i.e., cost per unit of resource utilized, and not regulating carbon emissions.    As a result, unlike \carbonContainerS, HotSpot will throttle containers if it is more cost-efficient to do so, and also does not employ vertical scaling since it is never cost-efficient to purchase resources and not use them.   Similarly, \carbonContainerS is also related to prior approaches to resource deflation~\cite{sharma2019resource,hpdc20} that vertically scale resources in response to cloud platforms reclaiming resources for high-priority tasks.   \carbonContainerS also ``inflate'' and ``deflate'' the resources allocated to a container but in response to changes in carbon emissions rather than scheduling decisions.

\vspace{0.07cm}
\noindent {\bf Carbon management.} There is substantial recent work on managing carbon emissions in cloud datacenters due to climate change~\cite{ecovisor,zero-carbon,gupta2021chasing,embodied2,patterson-2021,dean-carbon,arxiv-google,wait-awhile,cloudcarbon}.  

Some of this work has focused on embodied carbon~\cite{gupta2021chasing,embodied2}, which represents the carbon emissions from producing and using computing infrastructure.   While \carbonContainerS focuses on regulating operational carbon --- from powering servers --- its carbon metrics could be extended to include a server's amortized embodied carbon based on its expected lifetime and utilization.   In this case, amortized embodied carbon would increase as utilization decreases, since the server's total embodied carbon would be amortized over less computation.  While including amortized embodied carbon in our metric would be trivial and not significantly change \carbonContainerS' design or function, we explicitly did not include it because of multiple concerns: specifically, over whether server lifetime and embodied carbon can be accurately measured, and whether embodied carbon should be entirely attributed to cloud applications.  That is, since a cloud application's embodied carbon represents the manufacturer's operational carbon, current carbon accounting frameworks ``double count'' embodied carbon.  As a result, based on current carbon accounting frameworks, such as the GHG protocol~\cite{ghg}, combining embodied and operational carbon into a single metric may be misleading~\cite{hotcarbon-embodied,hotcarbon-hotair}. 

\vspace{0.03cm}
There has also been much recent work that has focused on optimizing operational carbon.  Much of this work advocates selecting datacenters that operate in regions with low-carbon energy~\cite{cloudcarbon,zero-carbon,patterson-2021,dean-carbon,arxiv-google}.  However, our analysis in \S\ref{sec:background} shows that there are few such regions.  Many workloads also cannot operate in these regions due to capacity limitations and latency constraints.  In addition, our analysis in \S\ref{sec:background} shows that dynamically migrating jobs to lower carbon regions is not beneficial due to both high migration overhead and a lack of opportunity, as regions' carbon-intensity rarely inverts.   We also compare \carbonContainerS with recent suspend/resume scheduling policies, such as Wait AWhile~\cite{wait-awhile}.   While suspend/resume scheduling is effective in reducing relative carbon emissions, it is only effective in regions with widely variable carbon-intensity, which only occurs when carbon-intensity is already low on average.  This approach is not effective in regulating carbon emissions in regions with high carbon-intensity, where it is most important, as they tend to have fewer carbon-intensity variations. 

\vspace{0.03cm}
Finally, \carbonContainerS differs from recent work that proposes ecovisors~\cite{ecovisor,enabling-socc21}, which virtualize the energy system and exposes visibility and control of it to applications.  Ecovisors burden applications with managing their own carbon emissions, and require application-specific modifications. In contrast, beyond setting the target carbon emissions rate, \carbonContainerS operate at the system-level, are entirely transparent to the application, and thus require no application-specific modifications.  That said, ecovisors have the flexibility to support \carbonContainerS, and we plan to implement \carbonContainerS on the ecovisor interface as future work.  \carbonContainerS represent one possible abstraction that ecovisors could support to make carbon management more transparent to applications. 

\section{Conclusion}
\label{sec:conclusion}
In this paper, we present the design and implementation of \carbonContainerS, a system-level facility for managing application-level carbon emissions.   \carbonContainerS enable applications to specify a maximum target carbon emissions rate, and then transparently enforce this rate via a combination of vertical scaling, migration, and suspend/resume while maximizing either a container's energy-efficiency or performance.  We motivated the need for \carbonContainerS by analyzing both energy's carbon-intensity and production workload characteristics and presented the design of \carbonContainerS' key mechanisms along with several policies. We evaluated \carbonContainerS using a prototype and in simulation using real workload traces. 
Our results show that \carbonContainerS are more effective than existing suspend/resume policies, i.e., they substantially increase performance while maintaining similar carbon emissions. Importantly, our approach is effective over a wide range of operating regimes, including geographic regions where carbon-intensity is high or variance is low. As future work, we plan to implement a range of higher-level policies using \carbonContainerS to demonstrate its efficacy for different types of compute and data-intensive applications.   
 
\vspace{0.07cm}
\noindent {\bf Acknowledgements.} This research is supported by NSF grants 2213636, 2136199, 2106299, 2102963, 2105494, 2021693, 2020888, 2045641, as well as VMware.

\balance
\bibliographystyle{plain}

\begin{thebibliography}{10}

  \bibitem{openai}
  Open{A}{I} {B}log, {A}{I} and {C}ompute.
  \newblock \url{https://openai.com/blog/ai-and-compute/}, March 16th 2018.
  
  \bibitem{azure-data}
  Azure {P}ublic {D}ataset.
  \newblock \url{https://github.com/Azure/AzurePublicDataset}, Accessed October
    2020.
  
  \bibitem{electricity-map}
  Electricity {M}ap.
  \newblock \url{https://www.electricitymap.org/map}, Accessed March 2022.
  
  \bibitem{google-pue}
  Google {D}ata {C}enters {E}fficiency.
  \newblock \url{google.com/about/datacenters/efficiency/}, Accessed March 2022.
  
  \bibitem{ghg}
  Greenhouse {G}as {P}rotocol.
  \newblock \url{https://ghgprotocol.org/}, Accessed March 2022.
  
  \bibitem{criu}
  Checkpoint/{R}estore in {U}serspace ({C}{R}{I}{U}).
  \newblock \url{https://criu.org/Main_Page}, Accessed June 2023.
  
  \bibitem{energy-proportional}
  Luiz~Andre Barroso and Urs H\"{o}lzle.
  \newblock The {C}ase for {E}nergy-{P}roportional {C}omputing.
  \newblock {\em Computer}, 40(12):33--37, December 2007.
  
  \bibitem{enabling-socc21}
  Noman Bashir, Tian Guo, Mohammad Hajiesmaili, David Irwin, Prashant Shenoy,
    Ramesh Sitaraman, Abel Souza, and Adam Wierman.
  \newblock Enabling {S}ustainable {C}louds: {T}he {C}ase for {V}irtualizing the
    {E}nergy {S}ystem.
  \newblock In {\em SoCC}, November 2021.
  
  \bibitem{hotcarbon-embodied}
  Noman Bashir, David Irwin, and Prashant Shenoy.
  \newblock On the {P}romise and {P}itfalls of {O}ptimizing {E}mbodied {C}arbon.
  \newblock In {\em Proceedings of the 2nd Workshop on Sustainable Computer
    Systems (HotCarbon)}, 2023.
  
  \bibitem{hotcarbon-hotair}
  Noman Bashir, David Irwin, Prashant Shenoy, and Abel Souza.
  \newblock Sustainable {C}omputing -- {W}ithout the {H}ot {A}ir.
  \newblock In {\em Proceedings of the First Workshop on Sustainable Computer
    Systems Design and Implementation (HotCarbon)}, 2022.
  
  \bibitem{zero-carbon}
  A.~Chien.
  \newblock Driving the {C}loud to {T}rue {Z}ero {C}arbon.
  \newblock {\em CACM}, 64(2), February 2021.
  
  \bibitem{xen-migration}
  Christopher Clark, Keir Fraser, Steven Hand, Jacob~Gorm Hansen†, Eric Jul†,
    Christian Limpach, Ian Pratt, and Andrew Warfield.
  \newblock Live {M}igration of {V}irtual {M}achines.
  \newblock In {\em NSDI}, April 2005.
  
  \bibitem{power-api}
  Maxime Colmant, Pascal Felber, Romain Rouvoy, and Lionel Seinturier.
  \newblock Watts{K}it: Software-{D}efined {P}ower {M}onitoring of {D}istributed
    {S}ystems.
  \newblock In {\em 17th IEEE/ACM International Symposium on Cluster, Cloud and
    Grid Computing (CCGRID)}, April 2017.
  
  \bibitem{azure-data-paper}
  Eli Cortez, Anand Bonde, Alexandre Muzio, Mark Russinovich, Marcus Fontoura,
    and Ricardo Bianchini.
  \newblock Resource {C}entral: {U}nderstanding and {P}redicting {W}orkloads for
    {I}mproved {R}esource {M}anagement in {L}arge {C}loud {P}latforms.
  \newblock In {\em Proceedings of the 26th Symposium on Operating Systems
    Principles}, SOSP '17, page 153–167, New York, NY, USA, 2017. ACM.
  
  \bibitem{cloudcarbon}
  Jesse Dodge, Taylor Prewitt, Remi Tachet~des Combes, Erika Odmark, Roy
    Schwartz, Emma Strubell, Alexandra~Sasha Luccioni, Noah~A. Smith, Nicole
    DeCario, and Will Buchanan.
  \newblock Measuring the carbon intensity of ai in cloud instances.
  \newblock In {\em 2022 ACM Conference on Fairness, Accountability, and
    Transparency}, FAccT '22, 2022.
  
  \bibitem{hpdc20}
  Alex Fuerst, Ahmed Ali-Eldin, Prashant Shenoy, and Prateek Sharma.
  \newblock Cloud-scale {V}{M}-deflation for {R}unning {I}nteractive
    {A}pplications on {T}ransient {S}ervers.
  \newblock In {\em ACM Symposium on High-Performance Parallel and Distributed
    Computing (HPDC)}, Stockholm, Sweden, June 2020.
  
  \bibitem{embodied2}
  Udit Gupta, Mariam Elgamal, Gage Hills, Gu-Yeon Wei, Hsien-Hsin~S. Lee, David
    Brooks, and Carole-Jean Wu.
  \newblock A{C}{T}: {D}esigning {S}ustainable {C}omputer {S}ystems with an
    {A}rchitectural {C}arbon {M}odeling {T}ool.
  \newblock In {\em ISCA}, June 2022.
  
  \bibitem{gupta2021chasing}
  Udit Gupta, Young~Geun Kim, Sylvia Lee, Jordan Tse, Hsien-Hsin~S Lee, Gu-Yeon
    Wei, David Brooks, and Carole-Jean Wu.
  \newblock Chasing {C}arbon: The {E}lusive {E}nvironmental {F}ootprint of
    {C}omputing.
  \newblock In {\em 2021 IEEE International Symposium on High-Performance
    Computer Architecture (HPCA)}. IEEE, 2021.
  
  \bibitem{cdn1}
  Vani Gupta, Prashant Shenoy, and Ramesh Sitaraman.
  \newblock Combining {R}enewable {S}olar and {O}pen {A}ir {C}ooling for
    {I}nternet-scale {D}istributed {N}etworks.
  \newblock In {\em e-Energy}, June 2019.
  
  \bibitem{rosing-isca}
  V.~Kontorinis, L.~Zhang, B.~Aksanli, J.~Sampson, H.~Homayoun, E.~Pettis,
    D.~Tullsen, and T.~Rosing.
  \newblock Managing {D}istributed {U}{P}{S} {E}nergy for {E}ffective {P}ower
    {C}apping in {D}ata {C}enters.
  \newblock In {\em ISCA}, June 2012.
  
  \bibitem{google-osdi20}
  Shaohong Li, Xi~Wang, Faria Kalim, Xiao Zhang, Sangeetha~Abdu Jyothi, Karan
    Grover, Vasileios Kontorinis, Nina Narodytska, Owolabi Legunsen, Sreekumar
    Kodakara, et~al.
  \newblock Thunderbolt: Throughput-{O}ptimized, {Q}uality-of-{S}ervice-{A}ware
    {P}ower {C}apping at {S}cale.
  \newblock In {\em USENIX Symposium on Operating System Design and
    Implementation (OSDI)}, November 2020.
  
  \bibitem{linux-containers}
  Canonical Ltd.
  \newblock Linux {C}ontainers.
  \newblock \url{https://linuxcontainers.org/}.
  
  \bibitem{masanet}
  Eric Masanet, Arman Shehabi, Nuoa Lei, Sarah Smith, and Jonathan Koomey.
  \newblock Recalibrating {G}lobal {D}ata {C}enter {E}nergy-use {E}stimates.
  \newblock {\em Science}, 367(6481):984--986, February 2020.
  
  \bibitem{arxiv-google}
  David Patterson, Joseph Gonzalez, Urs Hölzle, Quoc Le, Chen Liang,
    Lluis-Miquel Munguia, Daniel Rothchild, David So, Maud Texier, and Jeff Dean.
  \newblock The {C}arbon {F}ootprint of {M}achine {L}earning {T}raining {W}ill
    {P}lateau, {T}hen {S}hrink.
  \newblock Technical report, Google Inc., April 2022.
  
  \bibitem{dean-carbon}
  David Patterson, Joseph Gonzalez, Quoc Le, Chen Liang, Lluis-Miquel Munguia,
    Daniel Rothchild, David So, Maud Texier, and Jeff Dean.
  \newblock Carbon {E}missions and {L}arge {N}eural {N}etwork {T}raining.
  \newblock Technical report, arXiv, April 2021.
  
  \bibitem{patterson-2021}
  David Patterson, Joseph Gonzalez, Quoc Le, Chen Liang, Lluis-Miquel Munguia,
    Daniel Rothchild, David So, Maud Texier, and Jeff Dean.
  \newblock Carbon {E}missions and {L}arge {N}eural {N}etwork {T}raining, 2021.
  
  \bibitem{capping-asplos}
  Varun Sakalkar, Vasileios Kontorinis, David Landhuis, Shaohong Li, Darren
    De~Ronde, Thomas Blooming, Anand Ramesh, James Kennedy, Christopher Malone,
    Jimmy Clidaras, et~al.
  \newblock Data {C}enter {P}ower {O}versubscription with a {M}edium {V}oltage
    {P}ower {P}lane and {P}riority-{A}ware {C}apping.
  \newblock In {\em ACM Symposium on Architectural Support for Programming
    Languages and Operating Systems (ASPLOS)}, March 2020.
  
  \bibitem{sharma2019resource}
  Prateek Sharma, Ahmed Ali-Eldin, and Prashant Shenoy.
  \newblock Resource deflation: A new approach for transient resource
    reclamation.
  \newblock In {\em Proceedings of the Fourteenth EuroSys Conference 2019}, pages
    1--17, 2019.
  
  \bibitem{hotspot}
  Supreeth Shastri and David Irwin.
  \newblock Hot{S}pot: Automated {V}{M} {H}opping in {C}loud {S}pot {M}arkets.
  \newblock In {\em ACM Symposium on Cloud Computing (SoCC)}, Santa Clara,
    California, September 2017.
  
  \bibitem{power-containers}
  Kai Shen, Arrvindh Shriraman, Sandhya Dwarkadas, Xiao Zhang, and Zhuan Chen.
  \newblock Power {C}ontainers: An {O}{S} {F}acility for {F}ine-grained {P}ower
    and {E}nergy {M}anagement on {M}ulticore {S}ervers.
  \newblock In {\em ACM Conference on Architectural Support for Programming
    Languages and Operating Systems (ASPLOS)}, March 2013.
  
  \bibitem{followsun}
  Zhiming Shen, Qin Jia, Gur-Eyal Sela, Ben Rainero, Weijia Song, Robert van
    Renesse, and Hakim Weatherspoon.
  \newblock Follow the {S}un through the {C}louds: Application {M}igration for
    {G}eographically {S}hifting {W}orkloads.
  \newblock In {\em ACM Symposium on Cloud Computing (SoCC)}, Santa Clara,
    California, October 2016.
  
  \bibitem{ecovisor}
  Abel Souza, Noman Bashir, Jorge Murillo, Walid Hanafy, Qianlin Liang, David
    Irwin, and Prashant Shenoy.
  \newblock Ecovisor: A {V}irtual {E}nergy {S}ystem for {C}arbon-{E}fficient
    {A}pplications.
  \newblock In {\em ASPLOS}, March 2023.
  
  \bibitem{mccallum}
  Emma Strubell, Ananya Ganesh, and Andrew Mc{C}allum.
  \newblock Energy and {P}olicy {C}onsiderations for {M}odern {D}eep {L}earning
    {R}esearch.
  \newblock In {\em AAAI Conference on Artificial Intelligence (AAAI)}, February
    2020.
  
  \bibitem{wait-awhile}
  Philipp Wiesner, Ilja Behnke, Dominik Scheinert, Kordian Gontarska, and Lauritz
    Thamsen.
  \newblock Let's {W}ait {A}while: How {T}emporal {W}orkload {S}hifting {C}an
    {R}educe {C}arbon {E}missions in the {C}loud.
  \newblock In {\em Proceedings of the 22nd International Middleware Conference
    (Middleware)}, December 2021.
  
  \bibitem{wan-migration}
  Timothy Wood, K.K. Ramakrishnan, Prashant Shenoy, and Jacobus~Van der Merwe.
  \newblock Cloud{N}et: Dynamic {P}ooling of {C}loud {R}esources by {L}ive
    {W}{A}{N} {M}igration of {V}irtual {M}achines.
  \newblock In {\em International Conference on Virtual Execution Environments
    (VEE)}, Newport Beach, CA, March 2011.
  
  \end{thebibliography}

\end{document}